\documentclass[prl, twocolumn, showpacs]{revtex4}
\usepackage{graphicx,epsfig}
\usepackage{bm}
\usepackage{dcolumn}

\def\bea{\begin{eqnarray}}
\def\eea{\end{eqnarray}}
\def\be{\begin{equation}}
\def\ee{\end{equation}}

\begin{document}
\title{Inverse approach to atomistic modeling: Applications to a-Si:H and g-GeSe$_2$ }

\author{Parthapratim Biswas}
\email{Partha.Biswas@usm.edu, pbiswas@orca.st.usm.edu}
\affiliation{Department of Physics and Astronomy, The University of Southern Mississippi, 
Hattiesburg, MS 39401}

\author{D.~A.~Drabold}
\email{drabold@ohio.edu}
\affiliation{Department of Physics and Astronomy, Ohio University, Athens, OH 45701 }


\begin{abstract}
We discuss an inverse approach for atomistic modeling of glassy materials. The focus 
is on structural modeling and electronic properties of hydrogenated amorphous silicon 
and glassy GeSe$_2$ alloy. The work is based upon a new approach ``experimentally 
constrained molecular relaxation (ECMR)". Unlike conventional 
approaches (such as molecular dynamics (MD) and Monte Carlo simulations(MC), where 
a potential function is specified and the system evolves either deterministically 
(MD) or stochastically (MC), we develop a novel scheme to model structural configurations 
using experimental data in association with density functional calculations. We have 
applied this approach to model hydrogenated amorphous silicon  and glassy GeSe$_2$. The 
electronic and structural properties of these models are compared with experimental 
data and models obtained from conventional molecular dynamics simulation. 
\end{abstract}

\maketitle

\section{Introduction} 

In conventional electronic structure problem one starts with a formulation of a model 
that consists of a set of atomic or molecular coordinates and an interacting potential 
or Hamiltonian. The electronic density of states is obtained either by solving the 
Schr\"odinger equation within the first-principles density functional formalism or by 
constructing a semi-empirical Hamiltonian to compute the electronic 
eigenstates, total energy, and response functions to compare with experiments~\cite{Martin}.
For a crystalline system, such an approach is fairly straightforward (at least in principle), 
although in many cases it can be computationally very expensive. Disordered materials, 
however, pose the problem that is highly non-trivial due to lack of structural information 
about the materials. Amorphous materials are archetypal examples of this class, which are 
characterized by presence of short and medium range order ranging from few angstroms to several 
nanometers~\cite{elliott-book}. Since the position of an atom is not known in the amorphous state, 
one needs to proceed by simultaneously optimizing the geometry and calculating the electronic 
properties and structure for that geometry. While there exists a number of methods that can 
address this problem with a varying degree of accuracy (such as Car-Parrinello~\cite{cp-md} and 
first-principles molecular dynamics~\cite{Martin}), almost all cases where the problem is characterized by 
long time and large length scale, these techniques are computationally overkill and largely 
inapplicable for problems that require realistic modeling of complex amorphous 
systems. The so-called order-N methods (that scale linearly with system size N) can deal 
with large number of atoms; however, their use is often severely limited in structural modeling of 
amorphous materials due to lack of knowledge in spectral properties (e.g. the nature of 
density matrix and the presence of spectral gap).  First-principles molecular 
dynamics is, therefore, not practically viable for large scale modeling that are necessary to study 
some of the novel properties of materials associated  with amorphous state (such as medium range 
order and diffusion in glasses). The use of empirical potentials make it possible to run 
molecular dynamics simulation for several thousands atoms with a time scale of the order of 
several nanoseconds, but for many complex systems empirical potentials are not reliable enough 
to describe the geometry and local chemistry correctly. 

In this paper, we discuss applications of an alternative approach to model materials using a 
combination of experimental data and a suitable force-field (either first-principles, tight-binding 
or empirical). Instead of taking a direct approach, we take an inverse 
approach where experimental data are enforced to build atomic configurations that have the desired 
structural and electronic properties. Reverse Monte Carlo (RMC) is a classic example of 
such a method that has been discussed by several authors~\cite{McG, Biswas, Walters, Gere}. 
Here we briefly mention an extension of reverse Monte Carlo by merging experimental data 
using a suitable force-field known as Experimental Constrained Molecular Relaxation 
(ECMR), and discuss its application to hydrogenated amorphous silicon and glassy-GeSe$_2$.

\section{Experimentally constrained molecular relaxation (ECMR)}

The reverse Monte Carlo (RMC) method developed by McGreevy and colleagues is a classic 
example of an inverse problem in materials modeling~\cite{McG}. The method constructs a model 
configuration by making use of all available experimental information. 
The central idea is to set up a generalised cost function containing 
as much information as possible, and then to minimize the function for generating configurations toward 
exact agreement with one or more experiments. In order to reduce the solution space and 
to explore a limited part of physical subspace of configuration, a number of chemical, 
geometrical and topological constraints can be added. The mathematical structure 
of this problem is equivalent to the constraint optimization ``traveling salesman" 
problem: 
\be 
\label{eq-1}
\xi=\sum_{j=1}^{K} \sum_{i=1}^{M_K} \eta_{i}^j \{F^j_E(Q_i)- F^j_c(Q_i)\}^2
   + \sum_{l=1}^L \lambda_l P_l
\ee
\noindent
The coefficients 
${\eta_j}$ and ${\lambda_j}$ are the appropriate weight factors for each data set (F) and the 
constraints (P). The quantity $Q$ is the appropriate generalized variable associated with 
experimental data $F(Q)$. In order to prevent the atoms getting too close to each other, a 
certain cutoff distance is imposed, which is typically of the order of interatomic spacing. 
RMC has been applied successfully to variety of materials~\cite{McG}. However, the principal 
difficulty is the lack of uniqueness of the method. In absence of information from high 
order correlation functions, the method produces a range of configurations that are 
consistent with input experimental data, but may not be physically meaningful.  The imposition of
topological or chemical constraints can ameliorate the problem, but cannot eliminate 
completely.

The experimentally constrained molecular relaxation (ECMR) has been designed to overcome some of 
the problems above~\cite{ecmr-paper}. Instead of relying on experimental information and a set 
of constraints only, one additionally employs an approximate energy functional to describe the 
dynamics correctly by merging first-principles density functional (or semi-empirical) as well 
as experimental data. The purpose of the energy functional is to guide the system {\em approximately}
in the augmented configurational space defined by experimental data and other constraints. This 
largely reduces the number of unphysical solutions that are mathematically correct but fail to 
satisfy the dynamical behavior correctly. Introduction of an approximate energy functional constrains 
the system to evolve on a restricted {\em but} more realistic energy surface, and thereby accelerates 
to converge toward reliable structural configurations during the course of simulation.  
The configuration obtained from the method is not only a minimum (metastable or global) 
of an appropriate energy functional but also consistent with the input experimental information. 
Symbolically, we can write the grand penalty function in the ECMR as: 

\be 
\Xi({\bf Q}, {\bf r}) = \xi({\bf Q}) \; \oplus \; \gamma \; E({\bf r}) 
\label{eq-grand} 
\ee 

In equation (\ref{eq-grand}), the symbol $\oplus$ stands for direct sum of the configuration 
space of penalty function $\xi ({\bf Q})$ and that of the energy functional $E({\bf r})$~\cite{note}. 
In the limit $\gamma$ is infinitesimally small, the method reduces to a inverse method (RMC 
in the present case), whereas for a very large value of $\gamma$ the method is 
equivalent to a direct method of minimizing the total energy. 

\section{Applications: glassy-GeSe$_2$ and a-Si:H } 
\begin{figure}
\includegraphics[width=2.5 in, height=2.5 in, angle=-90]{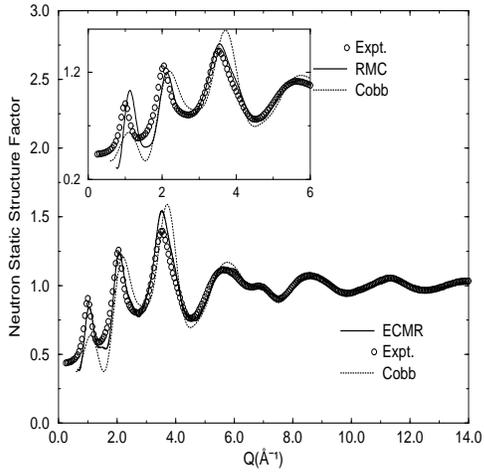}
\caption{ Neutron weighted static structure factors for the ECMR, reverse 
Monte Carlo (RMC) and decorate-and-relax (DR) model. The result for the DR 
model is taken from the Ref.~\onlinecite{Cobb}.  The low wave vector part 
is magnified in the inset for comparison to our model with experimental 
data and the DR model. 
}
\label{fig2}
\end{figure}

Amorphous GeSe$_2$ is a classic glass former and has interesting physical properties 
that are difficult to model via conventional molecular dynamics simulation~\cite{Cobb}. The material
strongly shows the presence of intermediate range order in the form of a 
first sharp diffraction peak (FSDP) in Neutron diffraction measurements. The origin of 
this intermediate range order is generally attributed to the presence of tetrahedral 
motifs having edge- and corner-sharing topology. Raman spectroscopy and Neutron diffraction 
provide useful information about the topological structure of the material~\cite{Jackson, Susman}.
The GeSe$_2$ model simulated in our ECMR work consists of 647 atoms.  The static structure 
factor (Neutron-weighted) of the final model is plotted in figure \ref{fig2} along with the 
experimental data and the data obtained by Cobb et al. in Ref.~\onlinecite{Cobb}.  It is 
clear from the figure that experimental data fit with the theoretical values obtained from 
ECMR model reasonably good. The structure at large Q matches quite well, but at small 
Q there exists deviations from the experimental peak position.  

\begin{figure}
\includegraphics[width=2.0 in, height=2.5 in, angle=-90]{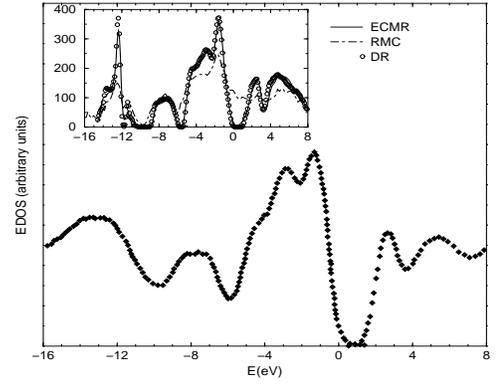}
\caption{ The experimental and computed electronic density of states (inset) 
of g-GeSe$_2$. The results for models obtained from the ECMR, reverse Monte Carlo (RMC), and 
decorate-and-relax (DR) are shown in inset. The results for the DR model are 
taken from Ref.~\onlinecite{Cobb} for comparison. 
}
\label{fig3}
\end{figure}

Since the network topology of g-GeSe$_2$ is very much influenced by the presence of 
edge- and corner-sharing tetrahedra, we have studied the presence of such tetrahedra 
in our model. As mentioned before, Raman and Neutron experiments~\cite{Jackson, Susman} 
revealed that about 33\% to 40\% of Ge atoms are involved in forming edge-sharing 
tetrahedra. The corresponding percentage in our model is found to be 38\%. This is 
particularly important in view of the fact that such information is not included 
in our starting ECMR model. It is remarkable to note that imposition of partial pair 
correlation functions and first-principles relaxation via ECMR does introduce the correct 
topological structure in the network. We have also observed that  81\% of Ge 
atoms in our model are 4-fold coordinated of which approximately 75\%  form predominant Ge-center  motifs 
$Ge(Se_{\frac{1}{2}})_4$ while  6\% are ethane-like $Ge_2(Se_{\frac{1}{2}})_6$ units.  
The remaining Ge atoms are 3-fold coordinated and are mostly found to be bonded as 
Ge--Se$_3$ units. The experimental radial distributions (partial) provide the partial 
nearest neighbor coordination number. The values for Ge-Ge, Ge-Se and Se-Se are given 
by 0.25, 0.20 and 3.7 respectively, whereas the corresponding values from our ECMR model
are 0.17, 0.30 and 3.68 respectively. 
In addition to studying the structural properties, we have also computed the electronic 
properties of the model. The electronic density of states provide an additional check 
and can be compared with X-ray photo emission data (XPS).  In figure \ref{fig3}, we have plotted 
the electronic density of states along with the experimental data. The electronic density 
of states of the ECMR model (shown in the inset) are in good agreement with the XPS data, 
and establishes further credibility of our method. 

We have also applied our ECMR approach to model hydrogenated amorphous silicon. For this 
purpose, we started with a pure amorphous silicon configuration obtained via Reverse Monte 
Carlo simulation. The particular RMC scheme to generate amoprhous silicon is discussed in 
Ref.~\onlinecite{rmc}. The initial size of the model is 500 atoms, which is then hydrogenated 
following a method similar to but not identical to Holender and Morgan~\cite{morgan}. In 
our scheme, the dangling bonds are passivated by introducing H atoms in the RMC-generated 
continuous random network. Once the dangling bonds are identified, passivation is done by 
placing a H atom at a distance of 1.45 {\AA} to 1.65 {\AA}. The atoms are placed along the 
direction vector opposite to the sum of the three vectors connecting the central atom and 
its three neighbors. Once the passivation is done, the density of the model is adjusted 
to experimental density and the system is subjected to ECMR iteration to make the resulting 
configuration consistent with both experimental data and the total energy. Any additional dangling 
bonds generated during ECMR iteration are also passivated, and the process is repeated until 
the dangling bonds cease to exist. The resulting model of hydrogenated amorphous silicon 
consists of 540 atoms with 7.4{\%} H atom in the network. The model has defect concentration 
of about 2{\%}, which consists of 5-fold coordinated floating bonds. 

\begin{figure}
\includegraphics[width=2.5 in, height=2.5 in, angle=0]{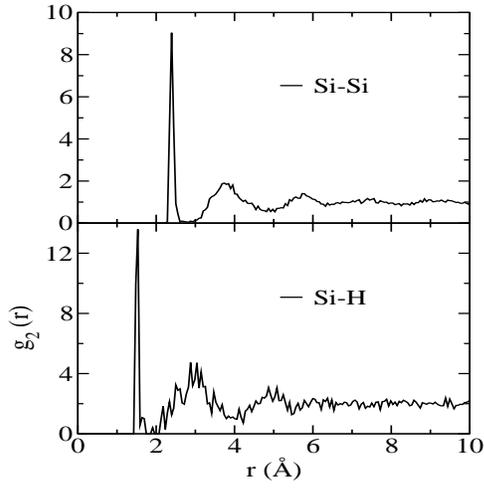}
\caption{The partial pair correlation functions for Si-Si (upper panel) 
and and Si-H (lower panel) atoms for the 540-atom ECMR model described in 
the text. 
}
\label{fig4}
\end{figure}

\begin{figure}
\includegraphics[width=2.5 in, height=2.5 in, angle=0]{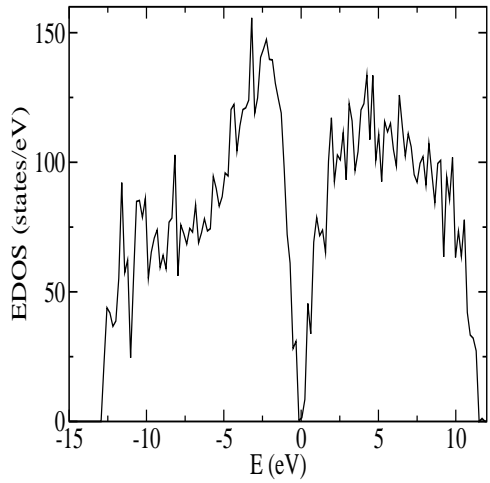}
\caption{ The electronic density of states of the 540-atom model of 
a-Si:H obtained from ECMR method within the local density 
approximation. The band gap is clearly visible in the spectrum along 
with a few tail states originated from the floating bonds present 
in the network } 
\label{fig5}
\end{figure}

The partial radial distribution functions for the model are plotted in figure \ref{fig4}. For the 
Si-Si case, most of the information are already included in the starting model, and hence 
we do not expect much changes in the final model. For Si-H part, however, it is important 
to note that during hydrogenation and subsequent relaxation of the model via ECMR, the 
system does not change much of the pair correlations between Si and H atoms. 
The electronic density of states (EDOS) for this 540-atom model is plotted in figure \ref{fig5}. The EDOS 
is obtained using the first-principles density functional code {\sc Siesta} within 
the local density approximation. The model clearly shows the presence of a gap in 
the spectrum along with a few tail states near the band edges. This is due to presence of few 
5-fold floating bonds in the network. As a further test of our model, we have also computed 
the vibrational density of states (VDOS) within the harmonic approximation by constructing 
the dynamical matrix using the electronic forces. The eigenvalues of the dynamical matrix 
are related to the square of the angular frequency of vibrations. The VDOS in figure \ref{fig5} 
distinctly show the presence of acoustic and optical peaks due to vibration of Si atoms. The 
high frequency vibrations in the density of states correspond to the H atom movement in the network. 

\section{Conclusion} 
We present an inverse approach to construct atomistic models of materials using a combination 
of experimental data and a suitable first-principles force-field. Using a generalised penalty 
function, we merge both the power of density functional theory with experimental data so that 
the configurations are not only a minimum (metastable or global) of an energy functional but also 
consistent with available experimental data. The method is applied to construct model 
configurations for two technologically important materials: glassy GeSe$_2$ and hydrogenated 
amorphous silicon. The structural, electronic and vibrational properties of the models are 
compared with experiments, and have been observed to be in very good agreement with experimental data.

\acknowledgments 
PB acknowledges the support of the University of Southern Mississippi under 
Grant No. DE00945. DAD thanks the US NSF for support under Grants DMR 0605890 
and 0600073.

\newpage 

\newpage 

\newpage 

\newpage 

\newpage 
\begin{figure}
\includegraphics[width=2.5 in, height=2.5 in, angle=0]{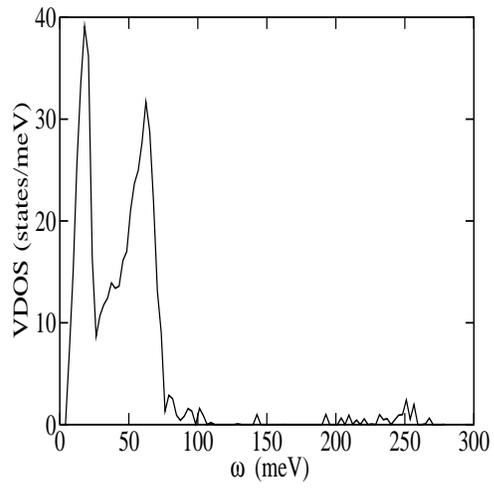}
\caption{ The vibrational density of the 540-atom model a-Si:H obtained 
within the harmonic approximation. The high frequency vibrations correspond 
to H atom movement in the network. The acoustic and optical peaks are clearly 
visible in the vibrational spectrum. 
}
\label{fig6}
\end{figure}

\begin{thebibliography}{99}
\bibitem{note} 
The grand penalty function can be constructed in many ways. If the functions are 
added algebraically, the symbol $\oplus$ stands for ordinary sum of two numbers.
\end{thebibliography}
\end{document}